\documentclass{article}
\usepackage{spconf,amsmath,graphicx,multirow,subcaption,xcolor,hyperref}

\usepackage{booktabs}
\usepackage{multirow}
\usepackage{tabularx}
\usepackage{enumerate}

\title{End-to-end lyrics Recognition with Voice to Singing   Style Transfer } 
\name{Sakya Basak$^\$$, Shrutina Agarwal$^\$$, Sriram Ganapathy$^\$$, Naoya Takahashi$^*$\thanks{This work was funded by Sony Corporation, Tokyo.}}
\address{$^\$$Learning and Extraction of Acoustic Patterns (LEAP) lab, Indian Institute of Science, Bangalore, India,\\
$^*$Sony Corporation, Tokyo, Japan.}
\graphicspath{{images/}}
\begin{document}
\ninept
\fontsize{9.0pt}{11.2pt}\selectfont

\maketitle
\begin{abstract}
Automatic transcription of  monophonic/polyphonic music is a challenging task due to the lack of availability of large amounts of transcribed data.  In this paper, we propose a data augmentation method that converts natural speech to singing voice based on vocoder based speech synthesizer. This approach, called voice to singing (V2S), performs the voice style conversion by modulating the F0 contour of the natural speech with that of a singing voice.  The V2S model based style transfer can generate good quality singing voice thereby enabling the conversion of large  corpora of natural speech to singing voice that is useful in building an E2E lyrics transcription system.  In our experiments on monophonic singing voice data, the V2S style transfer provides a significant gain (relative improvements of $21$ \%) for the E2E lyrics transcription system. We also discuss additional components like transfer learning and lyrics based language modeling to improve the performance of the lyrics transcription system.     
\end{abstract}
\section{Introduction}
\label{sec:intro}
In music, lyrics constitutes the textual component of singing voice. It forms an important constituent of the music signal that contributes to the emotional perception of the song \cite{ali2006songs} and also aids in foreign language learning \cite{good2015efficacy}. In music information extraction, two problems of interest are the   automatic alignment of lyrics and the automatic transcription of singing voice. The alignment problem is
the task of finding the timing of the word boundaries for the given lyrics with respect to the polyphonic audio~\cite{gupta2019acoustic}, while transcription is the task of recognizing the 
lyrics~\cite{mesaros2010automatic}. Several applications such as generating karaoke, music subtitling~\cite{dzhambazov2017knowledge},  query-by-singing \cite{hosoya2005lyrics}, keyword spotting, and automatic indexing of music according to transcribed keywords \cite{fujihara2008hyperlinking} rely on accurate alignment and transcription of music. 

The key challenges in the automatic recognition of lyrics is the unique style of singing voice, high variation of fundamental frequency and pronunciation, lack of large amounts of transcribed singing data and background scores. 
The earlier studies used a phoneme recognition approach \cite{gruhne2007phoneme}.        Mesaros et al.~\cite{mesaros2010automatic} adopted an automatic speech
recognition (ASR) based approach for phoneme and word recognition of singing vocals in monophonic and polyphonic music.
 In dealing with polyphonic music, one of the common approaches is to apply a voice source separation module \cite{fujihara2011lyricsynchronizer}.  In a recent work, Gupta et. al.  \cite{gupta2019acoustic}  attempted an adaptation of the models trained from solo music to polyphonic music for lyrics alignment. 
 
 In this paper, we attempt to perform automatic recognition of lyrics by utilizing the large amount of resources available for speech recognition. The conventional modular approach to ASR consists of several modules like acoustic model, lexical model and language model~\cite{hinton2012deep}.  The end-to-end (E2E) ASR is a simplified approach to overcome the limitations of the conventional approach by using a single neural network to perform an entirely data driven learning. 
It consists of a single deep neural network (DNN) model which is directly trained on words, sub-words or character targets, thereby  eliminating the need for the hand-crafted pronunciation dictionary. The earliest approach to E2E ASR used the connectionist temporal cost (CTC) function to optimize the
recurrent neural network model (RNN)~\cite{graves2014towards}.  The attention based models proposed recently  do 
not make any conditional independence assumption and they attempt
to learn an implicit language model using an encoder-decoder attention framework~\cite{bahdanau2016end}.   In order to combine the best of both worlds Watanabe et. al. proposed a hybrid CTC-attention model~\cite{watanabe2017hybrid}. In the last year, the performance of E2E models have been improved with the use of Transformer based architectures \cite{karita2019comparative}. 
The performance of the E2E ASR models can be further improved by using data augmentation techniques on the input features using  methods like time warping, frequency masking, and time masking~\cite{park2019specaugment}. 
However, E2E ASR tends to be data demanding in training, which makes it difficult to adopt this framework for lyrics transcription tasks as there is a considerable lack of large supervised  datasets.

In this paper, we propose a novel approach to data augmentation for end-to-end recognition of lyrics in singing voice. The proposed approach, termed as voice-to-singing (V2S), converts natural speech to singing voice using a vocoder based speech synthesizer \cite{WORLD}.  The V2S model uses the pitch contours from singing voice recordings along with the spectral envelope of the natural speech to perform voice to singing conversion. 
The proposed V2S approach can generate large amounts of ``singing'' voice for use in E2E model training. 
In addition, we investigate a transfer learning approach to leverage a large ``singing'' speech trained model.  The use of  source-separated singing voice data from polyphonic music, which is relatively easy to obtain compared to monophonic singing voice data, is also explored.
 We also develop a language model (LM) suitable for lyrics transcription by mining large text corpus of lyrics.
 The experiments are performed on the Dali corpus \cite{dali} and a proprietary music dataset provided by Sony. In these experiments, we show that the proposed V2S approach provides significant performance gains over baseline systems trained purely on natural speech.

The key contributions from this work are, 
{
\setlength{\leftmargini}{20pt} 
\begin{enumerate}[i]
\item We propose V2S approach for data augmentation to train the E2E lyric transcription model.
\item We investigate a transfer learning approach to leverage a large speech corpus and source-separated singing voice data from polyphonic music for E2E system. 
\item We develop a language model for lyrics transcription by mining large text corpus of lyrics. 
\item Experimental results on polyphonic and monophonic lyric transcription shows that the proposed V2S data augmentation, transfer learning using speech, source-separated singing voice data, speed perturbation and lyrics LM significantly improve the word error rate over the baseline system trained on natural speech.
\end{enumerate}
}
To the best of our knowledge, this paper constitutes one of the earliest efforts in developing E2E systems for automatic lyrics transcription of singing voice. 

\section{Related Prior Work}
Recently, Gupta et al. \cite{cgupta1} explored singing voice alignment on a cappella singing vocals using state of the art Google ASR model. The authors segmented the audio in $10$s segments, and the ASR model is used to get the time aligned transcription for the audio segment. The transcriptions are later refined using published lyrics and are used in training a conventional ASR. An iterative process is thereby ensued to refine the alignments. 
For polyphonic music, Sharma et al. \cite{bsharma}  developed a lyrics alignment system, where the ASR models were adapted to singing voices using speaker adaptive training on a small dataset of solo singing voices. In addition to this, the authors use source separated vocals from polyphonic music using convolutional neural network (CNN) based U-Net models. In a recent work on E2E models,  D.Stoller et al \cite{dstoller} investigated  Wave-U-Net  to predict character probabilities using raw audio. The short segment audio snippets ($10$s) are processed with the end-to-end model which uses CNN with CTC training. There has also been efforts  in exploring additional features like voicing, energy, auditory, spectral, and chroma during training of the model for alignment tasks~\cite{cgupta2}.

The background music may affect the intelligibility of the lyrics. Gupta et al. \cite{cgupta3} explored the significance of background music by performing genre based learning. Here, the music is divided into three different genres (hiphop, metal, and pop) and for each genre different phoneme and silence (non-vocal) models are trained using a standard HMM-GMM architecture. The authors report improvements using the genre specific modeling.

With the control of several acoustic features, Saitou et. al~\cite{saitou2007speech} proposed voice conversion approach to transform speech to singing style. 
An encoder-decoder approach to model-based learning of speech to singing voice conversion was explored  in Parekh~\cite{parekh2020speech}. 
However many of the past methods are not scalable to large volumes of speech data needed for acoustic model training in a E2E lyrics transcription system. 

\section{Proposed methods}
\label{sec:pagestyle}
In this section, we describe our approach on the V2S data augmentation, the transfer learning using source separated singing voice, and the language model development for lyrics.
\subsection{Voice to Singing (V2S)}
\label{sec:format}
The V2S converts natural speech to singing voice using a vocoder based speech synthesizer. We use the WORLD synthesizer \cite{WORLD} since it provide high-quality voice with low computational complexity. We first describe the WORLD model, followed by a description of the proposed V2S. 

\subsubsection{WORLD Decoder}
The speech analysis and synthesis systems which provide high quality reconstruction after modification of parameters are useful in various applications like singing synthesis and voice conversion systems. In order to process large amounts of data, the algorithms need to be computationally efficient.  The WORLD model is a fast algorithm which is vocoder based and generates high quality speech. The speech synthesis system consists of decomposing the speech signal into three components - 
fundamental frequency (F0)  (which is estimated using the DIO algorithm~\cite{morise2009fast}), spectral envelope (estimated using the CheapTrick algorithm  \cite{MORISE20151}) and  the excitation signal (estimated using the PLATINUM algorithm \cite{2012E1151} denoted as aperiodic parameter).  
The F0 information is used to determine the temporal positions of the origin of the vocal cord vibrations. 
The F0 information in WORLD vocoder is estimated using a  series of low-pass filters which gives multiple candidates. Using these mulitple candidate estimates for F0, a reliability measure based on the variance of the estimates is used to find the final  F0 value. 
\begin{figure}[t!]
    \centering
    \includegraphics[width=9cm, height=5.5cm]{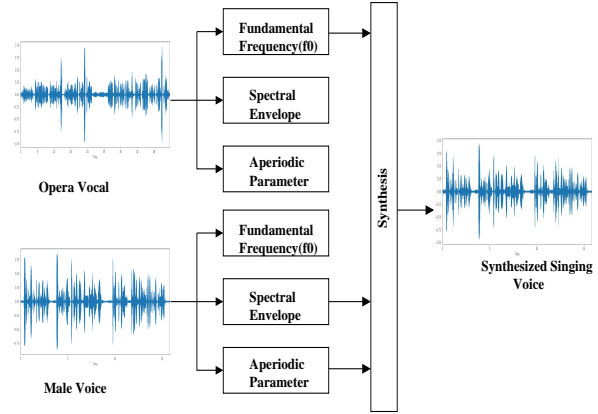}
    \caption{WORLD vocoder used in the proposed V2S system}
    \label{fig:WORLD_img}
    \vspace{-0.3cm}  
\end{figure}
\subsubsection {V2S using WORLD} 
The approach to data augmentation using the V2S model is shown in Figure~\ref{fig:WORLD_img}. We use the WORLD vocoder to independently decompose the natural speech and singing voice (like opera vocals) into the constituent components. The F0 contour from the singing voice is then used along with the spectral envelope and the aperiodic parameter from the natural speech and fed to the synthesizer.  The synthesized output is the singing voice version of the natural speech. 

The western opera vocal dataset consists of both male and female opera singers and during the synthesis we make sure that  the speech and the opera vocals are gender matched. Further, our analysis showed that, instead of randomly matching a  natural speech recording with an opera vocal sample, a technique for choosing the opera vocal based on the closest average F0 value with that of the speech signal under consideration improved the quality of the synthesized output drastically. To facilitate this operation, we perform the decomposition of the opera vocals in the dataset apriori and also store the average F0 value. Then, for the given speech signal under consideration, the decomposition is performed and the average F0 value is computed. The opera vocal that has the closest average F0 value from the database is chosen and its F0 contour is used in the synthesis of the singing speech. We did not perform any alignment of the F0 track with the speech signal.

\begin{figure}[t!]
\begin{subfigure}{0.4\textwidth}
\includegraphics[width=8cm,height=4cm]{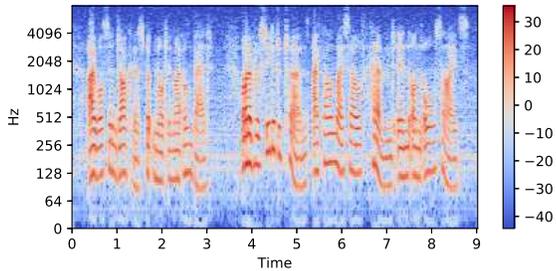} 
\vspace{-0.7cm}
\caption{Spectrogram of speech sample}
\label{fig:male_lib_spe_a}
\end{subfigure}
\begin{subfigure}{0.4\textwidth}
\includegraphics[width=8cm, height=4cm]{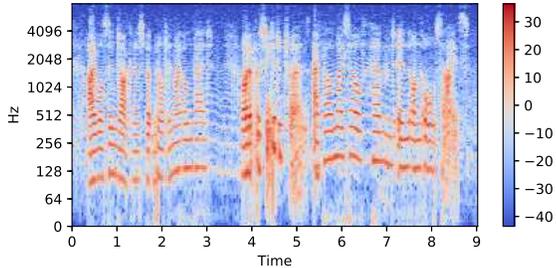}
\vspace{-0.6cm}
\caption{Spectrogram of synthesized waveform}
\label{fig:male_world_spe_b}
\end{subfigure}
\caption{ V2S  singing voice conversion of natural speech.}
\label{fig:spectrogram}
\vspace{-0.4cm} 
\end{figure}

An example illustration of the synthesized output from the proposed V2S approach is shown in Figure ~\ref{fig:male_world_spe_b}. The original speech sample is also show here in Figure ~\ref{fig:male_lib_spe_a}. As seen here, the synthesized output has different harmonicity properties. However, phonemic activity of the speech is  well preserved in the reconstructed output\footnote{Some audio samples  are available here - \url{https://github.com/iiscleap/V2S_Samples}}. 
\subsection{Transfer learning}
Although V2S can provide large amounts of synthetic singing voice data and provides significant performance gain over the model trained without V2S data augmentation as shown in Section \ref{sec:majhead}, there is still a  domain mismatch between the real and synthetic singing voice since the spectral features from standard speech are used in V2S. However, large amount of monophonic singing voice data is not available, especially for singing voice of professional artists. To overcome this challenge, we propose a transfer learning approach using source-separated singing voice data. Since polyphonic music with transcriptions are relatively easy to obtain like the DALI corpus \cite{dali}, we first separate a vocal track from polyphonic music using the state-of-the-art DNN-based source separation - D3Net \cite{Takahashi20D3net}. Then, E2E lyric recognition model trained with V2S data augmentation is fine tuned on the source-separated (SS) singing voice data. One can also consider fine tuning the E2E models on polyphonic music data directly. However, we experimentally show in Section \ref{sec:majhead} that incorporating the singing voice source separation works significantly better for polyphonic and monophonic cases. 

\subsection{Language model for lyrics text}
\label{sec:lm}
The probability distribution on words and their transition probability in lyrics are considerably different from standard speech. In particular,  lyrics often contain artistic expressions and may potentially violate many grammar usage rules in natural speech. To incorporate the difference, we develop a language model using a combination of text from audio-books and lyrics data from DALI \cite{dali}, Sony Singing Voice (SSV) dataset, and web resource\footnote{The lyrics data from www.lyrics.com}. This text data contained lyrics from a variety of genres like hip-hop, pop, classical, country-music, rock, jazz and consisted of about $5$M lines of text.


\begin{table}[t!]
    \centering
\caption{Performance of different E2E architectures trained on natural speech from Librispeech corpus. }
\begin{tabular}{|c|c|}
\toprule
Architecture  &  SSV \\ \hline
Hybrid LSTM \cite{watanabe2017hybrid} &  53.4 \\
Transformer \cite{karita2019comparative}  & 48.4\\
\hline
 \end{tabular}
    \label{tab:archDesign}
\end{table}

\begin{table}[t!]
\begin{center}
\caption{WERs of models trained with and without V2S data augmentation.
Here, SS-DALI denotes Source separated DALI, Poly-DALI denotes Polyphonic DALI. The results indicated as '--' correspond to conditions when some of the files fail to decode completely.}

\begin{tabular}{|p{3.0cm}|c c|c c|c|}
 \hline
  \multirow{2}{*}{Models}
  & \multicolumn{2}{c|}{SS-DALI}  &  \multicolumn{2}{c|}{Poly-DALI}  &   \multirow{2}{*}{SSV}  \\ 
  & Dev. & Test & Dev. & Test & \\ 
 \hline
 LS (original)  &  47.7 & 49.5 & 80.9 & 86.3 & 48.4 \\ 
 LS with V2S & 43.7& 46.8 & -- & -- &41.6  \\
  LS with V2S (F0 map) &\textbf{42.8}& \textbf{44.9} & \textbf{75.2} & \textbf{78.9} & \textbf{39.7} \\
 
 \hline
\end{tabular}
\label{tab:V2S}
\end{center}
\end{table}

\begin{table}[t!]
\centering 
\caption{Perplexity values on various test data for the LM. Here,  LS refers to audio-books in the LibriSpeech test and dev set. The LM model in the last row is refered to as the Lyrics LM.}
\vspace{-0.25cm}
\begin{center}
\begin{tabular}{|p{4.8cm}|p{0.7cm}|p{0.6cm}|p{0.7cm}|}
 \hline
 Training Data  &  LS  & SSV  & DALI  \\ 
 \hline
 Audio-book   & \textbf{28.7 }& 70.4 & 75.4 \\ 

---  fine-tuned on  SSV+DALI+web   & 449.8 & 57.6 & 66.3 \\
SSV+DALI+web   & 1002 & 75.9 & 81.6 \\ 
Audio-book+SSV+DALI+web    & 73.0 & \textbf{50.8} & \textbf{58.4} \\ 
 \hline
\end{tabular}
\label{tab:perplexity}
\end{center}
\vspace{-0.5cm} 
\end{table}




\begin{table*}[t!]
\begin{center}
\caption{WER values obtained after training the models on mapped modulated LibriSpeech and normal LibriSpeech. SS DALI - Source separated DALI, Poly DALI - Polyphonic DALI, SP - Speed Perturbation.}
\begin{tabular}{|p{4.3cm}|c c |c c |c ||c c|c c|c|}
 \hline
  \multirow{3}{*}{Model} & \multicolumn{5}{c||}{Audiobook LM} \vline  & \multicolumn{5}{c}{Lyrics LM} \vline \\\cline{2-11}
  & \multicolumn{2}{c|}{SS-DALI} & \multicolumn{2}{c|}{Poly-DALI} & \multirow{2}{*}{SSV} & \multicolumn{2}{c|}{SS-DALI} & \multicolumn{2}{c|}{Poly-DALI} & \multirow{2}{*}{SSV}\\
  &   Dev. &  Test & Dev. & Test &  &  Dev. &  Test & Dev. & Test &  \\ 
 \hline
 No fine tuning  & 61.4 & 65.3 & 84.4 & 87.3 & 39.7 & 58.9 & 62.5 & 82.2 & 85.3 & 38.1 \\ 
 
 Fine tuned on SS-DALI   & 46.2 & 49.4 & 77.3 & 80.9 & 39.1 & 44.8 & 47.0 & 75.2 & 78.1 & 38.4 \\ 
 
 Fine tuned on SS-DALI (SP) & 42.8 & 44.9 & 75.2 & 78.9 & 36.1 & \textbf{41.5} & \textbf{43.4} & 74.5 & 78.0 & \textbf{34.8}\\ 
  
 Fine tuned on Poly-DALI (SP)  & 56.0 & 58.7 & 59.1 & 61.8 & 45.5 & 55.1 & 57.4 &\textbf{57.9} & \textbf{60.4} & 44.8 \\ 
 \hline
\end{tabular}
\label{tab:SP} 
\end{center}
\vspace{-0.5cm}
\end{table*}

\section{Experimental Setup}
\label{sec:typestyle}
\subsection{Dataset}
We use the LibriSpeech corpus \cite{LibriSpeech} as the natural speech data corpus in all our experiments. 
The LibriSpeech corpus  contains $1000$  hours of read speech sampled at $16$ kHz. The training portion of the Librispeech contains $1129$ female speakers and $1210$ male speakers with each speaker providing $25$-$30$ minutes of speech.  The language model training data is also released as part of the Librispeech corpus which contains approximately $14,500$ public domain
books with around $800$M tokens in total and $900$K unique words. We use the train, test and dev partitions as prescribed in the release \cite{LibriSpeech}. 

For fine tuning to real singing voice, we use the DALI dataset \cite{dali}.
This dataset consists of  $1200$ English polyphonic songs with lyrics-transcription, which results in a total duration of $70$ hours. 
We split the dataset to a training, development, and test with a ratio of $8$:$1$:$1$. The splits are also carefully performed not to have any overlap in the artists performing in each of these splits. Further, the splits are also gender balanced to avoid any bias in training/testing. 
For testing on  monophonic singing voice to avoid the effect of accompaniment sounds, we also used a proprietary dataset from Sony Corp termed as Sony Singing Voice (SSV) dataset which consists of $88$ English Songs. 
Each of these songs have an approximate duration of $4$ min and had a mix of multiple genres. We have used $5$sec chunks of the audio for model training. The other model transformer model parameters follow the baseline Librispeech setup from ESPNET toolkit.  The singing voice in these recordings is  also of professional quality.  


\subsection{Language model}
We train two different language models (LM) - an audio book LM using the text resources from the Librispeech corpus and a lyrics LM  using  a combination of audio-book text with $5$M lines of lyrics text  described in Sec. \ref{sec:lm}.  The LMs are a recurrent neural network (RNN) model with $1$ layer of $1024$ LSTM cells.  The language model is incorporated in the E2E system as described in  ~\cite{bahdanau2016end}.  

\subsection{E2E ASR Framework}
We have used the ESPNET toolkit~\cite{espnet} to perform our E2E recognition experiments. The features used to train the architecture are log-filter bank features extracted using $30$ mel-spaced windows of duration $25$ms with a shift of $10$ms.   The E2E model used in most of the experiments is based on the Transformer architecture~\cite{karita2019comparative}.    The encoder used is a $12$-layer transformer  network with $2048$
 units in the projection
layer. The attention used is location attention and the decoder network is a $6$-layer Transformer network   with $2048$ units in the projection layer. During training, multiple cost functions are used~\cite{karita2019comparative}    like connectionist temporal cost (CTC) and the cross-entropy (CE) loss.  The model is trained using Adam optimization and training is performed for several epoch till the loss saturates on the validation data.  
The   
CTC-weight  is fixed at $0.3$ and during decoding the  beam-size is fixed at  $20$. Both the LMs (audio-book LM as well as the lyrics LM) have the same architecture and used $5000$ sub-word units at the output layer.

\section{Results}
\label{sec:majhead}
\textbf{E2E model architecture:} \hspace{1mm}
The first set of results shown in Table~\ref{tab:archDesign} highlights the lyrics transcription results on the SSV data using the Hybrid LSTM E2E architecture \cite{watanabe2017hybrid} and the transformer architecture \cite{karita2019comparative}. Both the models are trained on natural speech. As seen in this Table, the transformer architecture provides improved robustness to singing voice transcription task even when the model is trained for speech recognition task \cite{karita2019comparative}. 
\vspace{2mm}\\
\textbf{V2S:} \hspace{1mm}
Table~\ref{tab:V2S} shows the impact of the proposed V2S data augmentation for training the E2E model. The E2E model trained on natural speech  provides very high word-error-rates (WER) on polyphonic DALI dataset. The V2S improves the lyrics transcription performance on the SSV data significantly (average relative improvements of $14$\%  over the natural speech based WER). Further, the V2S applied with F0 mapping, where the opera vocals are selected to match the average F0 of the speech file under consideration, further improves the lyrics transcription accuracy. These results validate the effectiveness of proposed V2S data augmentation and F0 mapping (relative improvements of $18$ \%  on the SSV dataset). 
\vspace{2mm}\\
\textbf{LM comparison:} \hspace{1mm}
The comparison of various language models in terms of perplexity values is shown in Table~\ref{tab:perplexity}. The audio-book LM refers to the LM trained using the text data from the Librispeech corpus. The perplexity values highlighted here suggest that,  training the LM from the mixed corpus of speech and lyrics (from SSV, DALI and web) provides the best perplexity compared to either fine-tuning the audio-book LM or using only lyrics text for LM training. The best model (last row) in Table~\ref{tab:perplexity} is referred to as the lyrics LM in the experiments that follow. Further, the fine-tuned version of the audio-book LM (second row) is referred to as the audio-book LM for the remainder of the experiments. 
\vspace{2mm}\\
\textbf{Transfer learning:} \hspace{1mm}
The impact of fine-tuning the E2E models which were trained with V2S is shown in Table~\ref{tab:SP}. 
 These results show that fine-tuning on the source separated (SS) DALI training data can improve the performance on the DALI test data. Further, the application of speed perturbation in training \cite{ko2015audio} improves the lyrics transcription performance on both DALI and SSV test data. The application of V2S data augmentation in model training along with speed perturbation improves the lyrics transcription WER of SSV data relatively by about $28$ \% over the natural speech based E2E system. In addition, the WER results for the final system of $34.8$ \% WER on SSV data and $43.4$ \% on DALI data suggest that the highly challenging task of lyrics transcription on  monophonic/polyphonic music with large variety of music genres can  be explored with partial success using the  techniques proposed in this work.   
\section{Summary}
This paper presents a data augmentation method for E2E lyric transcription. The proposed method, termed as voice-to-singing (V2S), modulates the natural speech to a singing style by replacing a fundamental frequency contour of natural speech to that of singing voices using a vocoder based speech synthesizer. We also propose a transfer learning based approach to leverage a large amount of source-separated real singing voices from polyphonic music.
 The application of proposed methods are explored in the design of lyrics transcription system based on transformer based E2E model. Various experiments highlight the performance benefits of using the proposed V2S along with lyrics language modeling and transfer learning.  
 E2E ASR model is shown to provide useful features for other applications such as singing voice separation \cite{Takahashi20E2EASRSS}. Adopting the E2E lyric recognition model to such music applications is one of our future works.

 
  





\ninept
\bibliographystyle{IEEEbib}
\bibliography{ref}

\end{document}